\documentclass[12pt]{revtex4-1}%

\usepackage{epsfig}
\usepackage{psfrag}
\usepackage{amsmath}
\usepackage{amsfonts}
\usepackage{amssymb}

\newcommand{\be}{\begin{equation}}
\newcommand{\ee}{\end{equation}}
\newcommand{\bea}{\begin{eqnarray}}
\newcommand{\eea}{\end{eqnarray}}

\newcommand{\Q}{{\bf a}}

\begin{document}

\title{Cosmological solutions with  massive gravitons \\
in the bigravity theory }

\author{Mikhail~S.~Volkov }
\email[]{volkov@lmpt.univ-tours.fr}

\affiliation{ 
Laboratoire de Math\'{e}matiques et Physique Th\'{e}orique CNRS-UMR
6083, \\
Universit\'{e} de Tours, Parc de Grandmont, 37200 Tours, FRANCE}

\begin{abstract}

We present solutions describing homogeneous and 
isotropic cosmologies in the massive gravity theory with two dynamical metrics
recently proposed in arXiv:1109.3515 and claimed to be ghost free. 
These solutions can be spatially open, closed, or flat, and at early times 
they are sourced by the perfect fluid, 
while the graviton mass typically manifests itself 
at late times by giving rise to a cosmological term. 
In addition, there are also exotic solutions, for which already at  early times, when the matter 
density is high, the contribution of the graviton mass to the energy density
is negative and large enough to screen that of the matter contribution. 
The total energy can then be negative, which may result
in removing the initial singularity. For special parameter values there are also 
 solutions for which the 
two metrics effectively decouple and evolve independently of each other.  
In the limit where one of the gravitational coupling constant vanishes, 
such special solutions reduce to those found in arXiv:1107.5504
within the theory where one of the metrics is flat. 
  
\end{abstract}

\maketitle

\section{Introduction}

The currently observed acceleration of our universe \cite{Reiss} is the main
motivation of attempts to try to modify the theory of gravity, for example by
giving a tiny mass to the graviton. This can effectively give rise to a small
cosmological term leading to the late time acceleration \cite{DK}. The theory of massive 
gravity is not unique (see \cite{R} for a review) and there exist a number of its models, typically
parameterized by two metrics, one of which is physical, $g_{\mu\nu}$, 
while the other one is a flat reference metric $f_{\mu\nu}$. The coupling between the two is 
determined  by 
a scalar function of $g^{\mu\alpha}f_{\alpha\nu}$. 

Such models generically contain 
the negative norm ghost state in the spectrum \cite{BD}. 
There is, however, a particular model, we shall call it 
RGT model, that could be special, since it is the only one
that is ghost-free in the decoupling limit \cite{RGT}. In addition, this model 
was recently claimed to be ghost-free in the full theory \cite{HR1}. 
Even though the possibility to have a completely ghost-free massive gravity 
is sometimes disputed  \cite{CM}, the RGT model is certainly interesting. 
Its solutions describing spatially closed, open and flat homogeneous and isotropic
self-accelerating cosmologies were obtained in 
\cite{CV}, \cite{Mukohyama}. 

Quite recently, the generalizations of the the GRT model were proposed, first  
by promoting the reference metric $f_{\mu\nu}$ to be  
non-flat  \cite{HR2}, and next promoting it to be  dynamical \cite{HR3}, and 
it was argued that such generalized models remain  ghost-free. 
In what follows we shall study the cosmological solutions
within the bimetric theory of \cite{HR3}.

We find homogeneous and 
isotropic cosmologies which can be spatially open, closed, or flat.  
For the generic parameter values they can be of two main types. 
First of all, there are solutions for which the universe expansion 
is driven at early times by the ordinary matter, 
while the graviton mass manifests itself 
only at late times by giving rise to a cosmological term. 
In addition, there are also exotic solutions, for which already 
at  early times, when the matter 
density is high, the contribution of the graviton mass to the energy density
is large and negative and screens that of the matter contribution. 
The total energy can then be negative, which may result
in removing the initial singularity. In addition, for special  parameter values,
we find solutions 
for which the two metrics evolve independently of each other and
the physical metric shows the late time acceleration.  
In the limit where one of the gravitational coupling constant vanishes,
we shall call it the RGT limit,
they  reduce to those of  RGT theory found in  
\cite{CV}. 
It turns out that  the generic  solutions  do not reduce in the GRT limit
to any of the GRT theory, because both metrics remain then curved.

In what follows we present a rather detailed analysis of these  
solutions  as well as  their relation to the
GRT limit.  We address, in particular, the question (for some reason 
not discussed in the literature) of how to derive the Lagrangian field equations 
in the theory. This issue is in fact  not as trivial as it may seem, since 
the interaction between the two metrics is parameterized by
 $\gamma^\mu_{~\nu}=\sqrt{g^{\mu\alpha}f_{\alpha\nu}}$, whose direct variation is problematic, 
since the variation $\delta\gamma^\mu_{~\nu}$ does not commute in the matrix sense with 
$\gamma^\mu_{~\nu}$. 
We show how this problem 
can be handled withing the tetrad approach. 

The rest of the paper is organized as follows. In Sec.\ref{RGT} we describe the 
 bimetric generalization of the RGT theory, the tetrad approach, the variation  procedure
and the reduction to the spherically symmetric sector. Solution with the decoupled
metrics arising for special parameter values are described in Sec.\ref{decoupled},
while Sec.\ref{generic} presents a detailed analysis of the generic case.
Yet one more class of solutions, arising due to a different way to fulfill the conservation
condition, is described in Sec.\ref{non}, while the relation to the GRT limit is discussed
in Sec.\ref{eta}. We summarize our results in Sec.\ref{final}
and list in the Appendix 
the energy-momentum tensor components.

\section{The bimetric generalization of the RGT model \label{RGT}}

The theory is defined on  a four dimensional spacetime manifold ${\mathcal M}$ parameterized by 
coordinates $x^\mu$ and equipped with two metrics $g_{\mu\nu}(x)$ and  $f_{\mu\nu}(x)$.
Their kinetic terms are chosen to be of the 
standard Einstein-Hilbert form, with the corresponding couplings $G$ and $\eta G$.
The dynamics is 
determined  by the action  
\be
S=-\frac{1}{8\pi G}\,\int \left(\frac12\,R+m^2{\cal L}_{\rm int}\right)\sqrt{-g}\, d^4x 
-\frac{1}{16\pi \eta G}\,\int {\cal R}\sqrt{-f}\, d^4x 
+S_{\rm (m)}\,,
\ee
where $R$ and ${\cal R}$ are the Ricci scalars for $g_{\mu\nu}$ and $f_{\mu\nu}$, 
respectively, while $S_{\rm m}$ describes ordinary matter
(for example perfect fluid) which is supposed  to interact only with 
$g_{\mu\nu}$. 
The interaction between the two metrics  is defined by 
\bea                                           \label{lagr}
\mathcal{L}_{\rm int}=\frac{1}{2}(K^{2}-K_{\mu}^{\nu}K_{\nu}^{\mu})
+\frac{c_{3}}{3!}%
\,\epsilon_{\mu\nu\rho\sigma}
\epsilon^{\alpha\beta\gamma\sigma}K_{\alpha}^{\mu}
K_{\beta}^{\nu}K_{\gamma}^{\rho}+\frac{c_{4}}%
{4!}\,
\epsilon_{\mu\nu\rho\sigma}
\epsilon^{\alpha\beta\gamma\delta}K_{\alpha}^{\mu}
K_{\beta}^{\nu}K_{\gamma}^{\rho}K_{\delta}^{\sigma}\,,
\eea
with 
\be                                     \label{K}
K^\mu_\nu=\delta^\mu_\nu-\gamma^\mu_{~\nu}\,,
\ee
where $\gamma^\nu_{~\nu}$ is defined 
by the relation 
\be                                   \label{con}
\gamma^\mu_{~\sigma}\gamma^\sigma_{~\nu}=g^{\mu\sigma}f_{\sigma\nu}\,,
\ee
and $g^{\mu\nu}$ is the inverse of $g_{\mu\nu}$.  Apart from the 
gravitational coupling $G$, 
the theory contains three parameters $\eta,c_3,c_4$.  
In the limit where 
$\eta\to 0$ and $f_{\mu\nu}$ is flat it reduces to the RGT theory  \cite{RGT}.

The field equations are obtained by 
varying the action with respect to the metrics. 
A difficulty arises at this point, since varying the constraint \eqref{con}
gives  
\be
\delta\gamma^\mu_{~\sigma}\gamma^\sigma_{~\nu}
+\gamma^\mu_{~\sigma}\delta\gamma^\sigma_{~\nu}
=\delta g^{\mu\sigma}f_{\sigma\nu}+
g^{\mu\sigma}\delta f_{\sigma\nu}
\ee
and it is not obvious how to resolve this relation with respect to 
$\delta\gamma^\mu_{~\sigma}$. One could of course  consider $\gamma^\mu_{~\sigma}$
as independent variables and impose the constraint \eqref{con} within 
the Lagrange multiplier method. However, the Lagrange multiplier enters then 
the equations as an auxiliary field, and it is unclear how to get rid of it. 
Therefore, we adopt a different strategy, motivated by the approach of \cite{CM}, \cite{CV}.
 Let us introduce two  tetrads
$e_A^\mu$ and $\omega^A_\mu$ defined by the conditions 
\be                        \label{gf}
g^{\mu\nu}=\eta^{AB}e_A^\mu e_B^\nu\,,~~~~~~~~~
f_{\mu\nu}=\eta_{AB}\omega^A_\mu \omega^B_\nu\,,
\ee
where $\eta_{AB}={\rm diag}(1,-1,-1,-1)$ is the Minkowski metric. 
We shall also need their inverse  $e^A_\mu$ and $\omega_A^\mu$
such that $e_A^\mu e^B_\mu=\delta_A^B$ and $e_A^\mu e^A_\nu=\delta^\mu_\nu$,
similarly for $\omega_A^\mu$. 
Each of the tetrads $e_A^\mu$ and $\omega^A_\mu$ is defined up to 6 local 
Lorentz rotation, so that equations \eqref{gf} 
contain a 12-parameter gauge freedom. Let us now require that 
\be                       \label{const}
e_A^\mu\omega_{B\mu}=e_B^\mu\omega_{A\mu}\,,
\ee
where $\omega_{A\mu}=\eta_{AB}\omega^B_\mu$. 
This gives 6 local conditions, so that the two tetrads can now be Lorentz-rotated
only simultaneously, which leaves only the 6-parameter freedom of local rotations
in their definition. 
We then have a simple relation
\be                      \label{gam}
\gamma^\mu_{~\nu}=e^\mu_A\omega^A_\nu\,,
\ee
because 
\be                                   \label{con1}
\gamma^\mu_{~\sigma}\gamma^\sigma_{~\nu}
=e^\mu_A\omega^A_\sigma e^\sigma_B\omega^B_\nu
=e^\mu_A e^{A\sigma}\omega_{B\sigma}\omega^B_\nu=
g^{\mu\sigma}f_{\sigma\nu}\,,
\ee
which reproduces Eq.\eqref{con}. 
As a result, we can vary the action with respect to 
 $e_A^\mu$ and $\omega^A_\mu$ and the variation of $\gamma^\mu_{~\nu}$ will be 
obtained by varying Eq.\eqref{gam}. In order to take into account the 
conditions \eqref{const}, we add to the Lagrangian the term
\be
(e_A^\mu\omega_{B\mu}-e_B^\mu\omega_{A\mu})\lambda^{AB}
\ee 
where $\lambda^{AB}=-\lambda^{BA}$ are the 6 Lagrange multiplies. 
This allows us to consider $e_A^\mu$ and 
$\omega^A_\mu$ as independent variables.
Performing then the variation and expressing $\lambda^{AB}$ 
from the resulting equations,
we find that $\lambda^{AB}=+\lambda^{BA}$. 
Therefore, $\lambda^{AB}$ should be at the same time 
symmetric and antisymmetric on-shell,
so that it should vanish. 

As a result, we arrive at the following field equations
\begin{align}
R^\rho_\lambda-\frac12\,R\,\delta^\rho_\lambda&=m^2 T^\rho_\lambda
+8\pi G T^{{\rm (m)}\,\rho}_{~~~~\lambda} \,, \label{e1} \\
{\cal R}^\rho_\lambda-\frac12\,{\cal R}\,\delta^\rho_\lambda&=
\eta\,m^2 {\cal T}^\rho_\lambda    \,,        \label{e2}
\end{align}
where 
\be                                   \label{TTT}
T^\rho_\lambda=\tau^\rho_\lambda
-\delta^\rho_\lambda\,{\cal L}_{\rm int}\,,~~~~~~~
{\cal T}^\rho_\lambda=-\frac{\sqrt{-g}}{\sqrt{-f}}\, 
\tau^\rho_\lambda\,,
\ee
with 
\bea				
\tau^\rho_\lambda&=&e^\rho_B\frac{\partial {\cal L}_{\rm int}  }{\partial e_B^\lambda}
=\omega_\lambda^B\frac{\partial {\cal L}_{\rm int}  }{\partial \omega^B_\rho}= \notag \\
&=&(\gamma^\sigma_\sigma-3)\gamma^\rho_\lambda-
\gamma^\rho_\sigma\gamma^\sigma_\lambda -\frac{c_{3}}{2}%
\,\epsilon_{\lambda\mu\nu\sigma}
\epsilon^{\alpha\beta\gamma\sigma}\gamma_{\alpha}^{\rho}
K_{\beta}^{\mu}K_{\gamma}^{\nu}
-\frac{c_{4}}{6}\,
\epsilon_{\lambda\mu\nu\sigma}
\epsilon^{\alpha\beta\gamma\delta}
\gamma_{\alpha}^{\rho}
K_{\beta}^{\mu}K_{\gamma}^{\nu}K_{\delta}^{\sigma}\,.  \label{tau}
\eea
The Bianchi identities for the left-hand side of Eq.\eqref{e1} imply
the conservation condition 
\be                                   \label{T1} 
\stackrel{(g)}{\nabla}_\rho T^\rho_\lambda=0\,,
\ee
where $\stackrel{(g)}{\nabla}_\rho$ is the covariant derivative with respect to  
$g_{\mu\nu}$. It is worth noting that 
the matter energy-momentum tensor is conserved independently, in view 
of the diffeomorphism-invariance of the matter action $S_{\rm (m)}$,
\be                          \label{T2}
\stackrel{(g)}{\nabla}_\rho T^{{\rm (m)}\rho}_{~~~~\lambda}=0\,.
\ee
The Bianchi identities for the left-hand side of Eq.\eqref{e2} imply
that $\stackrel{(f)}{\nabla}_\rho {\cal T}^\rho_\lambda=0$, but these conditions 
in fact follow from Eq.\eqref{T1}, 
in view 
of the diffeomorphism-invariance of the interaction term 
$S_{\rm int}=\int {\cal L}_{\rm int}\sqrt{-g}\,d^4x$. 
Indeed, let us consider a diffeomorphism induced  by a vector field $\xi^\mu$. 
It induces variations
\bea
\delta e_A^\mu&=&e_A^\sigma\partial_\sigma\xi^\mu-\xi^\sigma\partial_\sigma e_A^\mu=
e_A^\sigma\stackrel{(g)}{\nabla}_\sigma\xi^\mu-
\xi^\sigma\stackrel{(g)}{\nabla}_\sigma e_A^\mu \,,\notag \\
\delta \omega^A_\mu&=&
-\partial_\mu\xi^\sigma\omega_\sigma^A
-\xi^\sigma\partial_\sigma\omega_\mu^A=
-\stackrel{(f)}{\nabla}_\mu\xi^\sigma\omega_\sigma^A
-\xi^\sigma\stackrel{(f)}{\nabla}_\sigma\omega_\mu^A,
\eea
while 
\bea
0&\equiv&\delta S_{\rm int}= \int \left\{
\frac{\partial ({\cal L}_{\rm int} \sqrt{-g}) }{\partial e_A^\mu}\,\delta e_A^\mu+
\frac{\partial ({\cal L}_{\rm int} \sqrt{-g})}{\partial \omega^A_\mu}\,\delta \omega^A_\mu
\right\}d^4x\notag \\
&=&-\int  \xi^\mu \stackrel{(g)}{\nabla}_\sigma T^\sigma_\mu
\sqrt{-g}\,d^4x
-\int \xi^\mu \stackrel{(f)}{\nabla}_\sigma {\cal T}^\sigma_\mu 
\sqrt{-f}\,d^4x. 
\eea
Since $\xi^\mu$ can be arbitrary, it follows that the conditions 
$\stackrel{(g)}{\nabla}_\rho T^\rho_\lambda=0$ imply that 
$\stackrel{(f)}{\nabla}_\rho {\cal T}^\rho_\lambda=0$.

If   $\eta\to 0$ 
then the source term in \eqref{e2} vanishes and one obtains ${\cal R}_{\mu\nu}=0$,
whose solution can be chosen to be flat metric, which can be written 
as $f_{\mu\nu}=\eta_{AB}\partial_\mu\Phi^A\partial_\nu\Phi^B$,
where $\Phi^A$ are sometimes called Stuckelberg fields. 
All the above considerations then still apply, in particular the tetrad formalism,
where it is sufficient to choose $\omega^A_\mu=\partial_\mu\Phi^A$. 
The basic field equations are then \eqref{e1},\eqref{T1},\eqref{T2} which determine 
 $g_{\mu\nu}$ and $\Phi^A$.

Let us return to the generic case with  $\eta\neq 0$.
In what follows we shall be considering solutions of equations \eqref{e1}, \eqref{e2},  
\eqref{T1}, \eqref{T2} with spherical symmetry. 
We introduce spherical coordinates 
$x^\mu=(t,r,\vartheta,\varphi)$ 
 and choose the tetrads to be  
\bea
e_0=\frac{1}{S}\,\frac{\partial}{\partial t }+q\,\frac{\partial}{\partial r },~~~~
e_1=p\,\frac{\partial}{\partial t }+\frac{1}{N}\,\frac{\partial}{\partial r },~~~~
e_2=\frac{1}{R}\,\frac{\partial}{\partial \theta },~~~~
e_3=\frac{1}{R\sin\vartheta}\,\frac{\partial}{\partial \varphi },~~~~
\eea
and 
\bea
\omega^0=a\,dt+c\,dr,~~~~\omega^1=d\,dt+b\,dr,
~~~~\omega^2=Ud\vartheta,
~~~~\omega^3=U\sin\vartheta d\varphi\,,
\eea 
where $S,q,N,p,R,a,b,c,d,U$ are functions of $t,r$. 
This implies the spherical symmetry for both metrics,
while using the residual diffeomorphisms in the $t,r$ subspace
one can always make the metric $g^{\mu\nu}$ diagonal.  
We therefore set 
\be
g^{0r}=e_0^0e_0^r-e_1^0e_1^r=\frac{q}{S}-\frac{p}{N}=0\,,
\ee
so that $q=pS/N$. Next, we consider the symmetry conditions \eqref{const},
of which the only non-trivial one is 
\be
-e_0^\mu\omega_{1\mu}+e_1^\mu\omega_{0\mu}=
e_0^\mu\omega^1_{\mu}+e_1^\mu\omega^0_{\mu}=
\frac{d}{S}+qb+ap+\frac{c}{N}=0\,,
\ee
from where $d=-apS-S^2pb/N-Sc/N$. 
We then notice that changing the parameter $p$ 
corresponds to the simultaneous local Lorentz rotations
of the two tetrads and does not change the metrics. 
We can therefore impose the gauge condition $p=0$, which finally gives 
the following most general expressions for the tetrads:
\bea               
e_0&=&\frac{1}{S}\,\frac{\partial}{\partial t },~~~~~~~~~
e_1=\frac{1}{N}\,\frac{\partial}{\partial r },~~~~~~~~~
e_2=\frac{1}{R}\,\frac{\partial}{\partial \theta },~~~~~~~~~
e_3=\frac{1}{R\sin\vartheta}\,\frac{\partial}{\partial \varphi },~~~~\notag \\
\omega^0&=&a\,dt+c\,dr,~~~~\omega^1=-\frac{cS}{N}\,dt+b\,dr,
~~~~\omega^2=Ud\vartheta,
~~~~\omega^3=U\sin\vartheta d\varphi\,.                    \label{tetrad}
\eea 
The corresponding metrics read 
\be
g_{\mu\nu}dx^\mu dx^\nu=S^2dt^2-N^2 dr^2-R^2(d\vartheta^2+\sin^2\vartheta d\varphi^2)
\ee
and
\be
f_{\mu\nu}dx^\mu dx^\nu=(a^2-\frac{S^2 c^2}{N^2})\,dt^2+2\frac{c(aN+Sb)}{N}\,dtdr
-(b^2-c^2)\, dr^2-U^2(d\vartheta^2+\sin^2\vartheta d\varphi^2),
\ee
while 
\be                                  \label{gamma}
\gamma^\mu_{~\nu}=e_A^\mu \omega^A_\nu=\left(
\begin{array}{cccc}
{a}/{S} & {c}/{S} & 0 & 0 \\
-{cS}/{N^2} & {b}/{N} & 0 & 0 \\
0 & 0 & {U}/{R} & 0 \\
0 & 0 & 0 & {U}/{R}
\end{array}
\right)\,,
\ee
and it is easy to verify that 
$\gamma^\mu_{~\sigma}\gamma^\sigma_{~\nu}
=g^{\mu\sigma}f_{\sigma\nu}$. We also notice that 
\be                           \label{det}
\frac{\sqrt{-g}}{\sqrt{-f}}
=\frac{1}{|e_A^\mu||\omega^A_\mu|}=
\frac{R^2}{U^2}\left(\frac{ab}{SN}+\frac{c^2}{N^2} \right)^{-1}\,.
\ee

We can now compute ${\cal L}_{\rm int}$ and the tensor $\tau^\mu_\nu$ defined by \eqref{tau}, 
they are 
 shown in the Appendix. Since  
our fields are only SO(3)-invariant, we have at the time being
$\tau^0_r\neq 0$,
$\tau^r_r-\tau^\vartheta_\vartheta\neq 0$. 
Our aim is to find homogeneous and isotropic solutions for $g_{\mu\nu}$\,,
in which case one should have $T^0_r=\tau^0_r=0$ and 
$T^r_r-T^\vartheta_\vartheta=\tau^r_r-\tau^\vartheta_\vartheta=0$. 
We therefore proceed to eliminate the components 
$\tau^0_r$ and $\tau^r_r-\tau^\vartheta_\vartheta$. 
One has 
\bea                                 \label{T0r}
\tau^0_r=\frac{c}{R^2S}
\left( -R\,(3R-2U)+ c_3\,(3R-U)(R-U)
+c_4\,(R-U)^2 \right).
\eea
For this to vanish, we can either choose $c=0$, or set to zero 
the expression between the parenthesis.

\section{Solutions with decoupled metrics \label{decoupled} }

Let us first consider the case where $c\neq 0$ and choose $U=CR$, 
where $C$ is a constant. Eq.\eqref{T0r} then becomes  
\be
\tau^0_r=
\frac{c}{S}\,\{2C-3+c_3(C^2-4C+3)+c_4(C-1)^2\},
\ee
which can be set to zero by adjusting the value of $c_4$,
but then  one finds 
\be
\tau^r_r-\tau^\vartheta_\vartheta=\frac{(C-1)c_3-C+2}{C-1}
\left(
C^2-\frac{Ca}{S}-\frac{Cb}{N}+\frac{c^2}{N^2}+\frac{ab}{SN}
\right),
\ee
which can in turn be set to zero by adjusting $c_3$. 
It follows that setting
\be
c_3=\frac{C-2}{C-1},~~~~~~c_4=-\frac{C^2-3C+3}{(C-1)^2},
\ee
so that $c_3(c_3-1)+c_4+1=0$, 
one achieves both $\tau^0_r=0$ and 
$\tau^r_r=\tau^\vartheta_\vartheta$. The $\tau^\mu_\nu$ components 
shown in the Appendix 
then reduce to 
\be
\tau^\mu_\nu=C(C-1)\left(\frac{c^2}{N^2}+\frac{ab}{NS}\right)\delta^\mu_\nu\,,
\ee
while
\be
{\cal L}_{\rm int}=C(C-1)\left(\frac{c^2}{N^2}+\frac{ab}{NS}-\frac{1}{C}   \right).
\ee
This gives
\be
T^\mu_\nu=\tau^\mu_\nu-\delta^\mu_\nu\,{\cal L}_{\rm int}=(C-1)\delta^\mu_\nu\,,
\ee
whereas using \eqref{det} 
\be
{\cal T}^\mu_\nu=-\frac{\sqrt{-g}}{\sqrt{-f}}\, 
\tau^\mu_\nu=\frac{1-C}{C}\,\delta^\mu_\nu\,.
\ee
The field equations \eqref{e1},\eqref{e2} therefore become 
\begin{align}
G^\mu_\nu&=m^2 (C-1)\delta^\mu_\nu
+8\pi G T^{{\rm (m)}\,\mu}_{~~~~\nu} \,, \label{ee1} \\
{\cal G}^\mu_\nu&=
\eta\,m^2 \frac{1-C}{C}\,\delta^\mu_\nu    \,,        \label{ee2}
\end{align}
so that the  equations for $g_{\mu\nu}$ decouple from those for $f_{\mu\nu}$. 
It is now easy to get cosmological solutions. Setting
\be
S=\Q(t),~~~N=\frac{\Q(t)}{\sqrt{1-kr^2}},~~~R=r\Q(t)
\ee
with $k=0,\pm 1$, so that 
\be                          \label{a0}
g_{\mu\nu}dx^\mu dx^\nu=\Q^2(t)\left(dt^2-\frac{dr^2}{1-kr^2}
-r^2(d\vartheta^2+\sin^2\vartheta d\varphi^2)\right)
\ee
and choosing $8\pi GT^{{\rm (m)}\,\mu}_{~~~~\nu}={\rm diag}(\rho(t),-P(t),-P(t),-P(t))$,
equations \eqref{ee1} reduce to 
\be                                \label{a1}
3\,\frac{\dot{\Q}^2+k\,\Q^2}{\Q^4}=m^2(C-1)+\rho\,,
\ee
where $\rho(t)$ is defined by the conservation condition 
\be                           \label{fluid}
\dot{\rho}+3\,\frac{\dot{\Q}}{\Q}\,(\rho+P)=0.
\ee
These equations describes the late time cosmological acceleration. 
If $\rho=\gamma P$ then $\rho\sim \Q^{-3-3/\gamma}$ so that for large $\Q$ the second term on
the right in \eqref{a1} becomes negligible. The dynamic is then driven by the
cosmological term $m^{2}(C-1)$, which we assume to be positive, so that $C>1$. 

It is worth noting that Eq.\eqref{a1} is exactly the same as Eq.(18) of \cite{CV}
obtained in the RGT theory. 
These 
solutions therefore  do not 
depend on weather the metric $f_{\mu\nu}$ is dynamical or not, which is due to 
the fact that equations \eqref{ee1} for $g_{\mu\nu}$ completely decouple from
equations \eqref{ee2} for $f_{\mu\nu}$. In order to solve equations \eqref{ee2}
 for $f_{\mu\nu}$ 
we notice that its components $f_{\vartheta\vartheta}=U^2$ and  
$f_{\varphi\varphi}=U^2\sin^2\vartheta$ are already fixed, since $U=Cr\Q(t)$, 
but $f_{00}$, $f_{0r}$, $f_{rr}$ are still free, 
since they contain three up to now
unspecified functions $a,b,c$. To see that this freedom is enough to fulfill the ten
equations \eqref{ee2}, we notice that one can consider $U$ as the new
radial coordinate. The time coordinate should also be changed, so that
\be
t\to T(t,r),~~~r\to U(t,r),
\ee
and the metric becomes 
\be                              \label{AdS1}
f_{\mu\nu}dx^\mu dx^\nu=f_{TT}\, dT^2+2f_{TU}dTdU+f_{UU}dU^2
-U^2(d\vartheta^2+\sin^2\vartheta d\varphi^2)\,,
\ee
where $f_{TT},f_{TU},f_{UU}$ are functions of $T,U$. 
The structure of the source term in \eqref{ee2} does not change in new coordinates, so that 
we should solve the Einstein equations with the negative cosmological term 
$\eta\,m^2 \frac{1-C}{C}$ to find a metric parameterized by the radial Schwarzschild
coordinate $U$. The solution is the anti-de Sitter metric 
\be                              \label{AdS}
f_{\mu\nu}dx^\mu dx^\nu=F^2\, dT^2-\frac{dU^2}{F^2}
-U^2(d\vartheta^2+\sin^2\vartheta d\varphi^2)\,,
\ee
where  $F^2(U)=1+\eta\,m^2 \frac{C-1}{3C}\,U^2$. One can now establish 
the relation to the $t,r$ coordinates, since we can read off the tetrad components
from \eqref{AdS}, but on the other hand they are given by \eqref{tetrad},
so that one can compare to obtain
\bea
\omega^0&=&FdT=F\dot{T}dt+FT^\prime dr=a\,dt+c\,dr     \notag \\
\omega^1&=&\frac{dU}{F}=\frac{C}{F}\,(\Q dr+r\dot{\Q}dt )=-c\sqrt{1-kr^2}\,dt+b\,dr\,.
\eea
This determines 
\be
b=\frac{C\Q}{F},~~~c=-\frac{Cr\dot{\Q}}{F\sqrt{1-kr^2}  },~~~~~a=F\dot{T}\,,
\ee
and also
\be                         \label{TTT1}
T=-\int \frac{Cr\dot{\Q}}{F^2\sqrt{1-kr^2}}\, dr\,.
\ee
Together with $U=Cr\Q(t)$, this establishes the correspondence 
between the $t,r$ and $T,U$ coordinates and also  specifies all 
the unknown functions in the solution.

\section{Generic solutions \label{generic}}

Let us now return to Eq.\eqref{T0r} with arbitrary $c_3,c_4$ and set $c=0$.  
This gives
$\tau^0_r=0$, while
$$
\tau^r_r-\tau^\vartheta_\vartheta=\frac{bR-UN }{NSR^2}\,
\{ US-3RS+aR+c_3(a-2US+3RS-2aR )+c_4(-US+Ua+RS-aR  )  \}.
$$
We now choose both metrics to be homogeneous and isotropic,
\be
S=\Q(t),~~N=\frac{\Q(t)}{\sqrt{1-kr^2}},~~R=r\Q(t),~~
a=\alpha(t),~~b=\frac{\beta(t)}{\sqrt{1-kr^2}},~~U=r\beta(t).
\ee
This insures that the energy-momentum tensors depend only on time 
and have the diagonal structure, 
$T^\mu_\nu={\rm diag}(T^0_0,T^r_r,T^r_r,T^r_r)$
and 
${\cal T}^\mu_\nu={\rm diag}({\cal T}^0_0,{\cal T}^r_r,{\cal T}^r_r,{\cal T}^r_r)$
(the explicit form of the tensor components can be read off from the formulas 
given in the Appendix). 
The independent equations are then the two Einstein equations 
\be                              \label{G}
G^0_0=m^2T^0_0+\rho,~~~~~{\cal G}^0_0=\eta m^2{\cal T}^0_0\,,
\ee 
as well as the conservation condition for ${T}^\mu_\nu$\,,
\be                                \label{T}
\dot{T^0_0}+3\,\frac{\dot{\Q}}{\Q}\,(T^0_0-T^r_r)=0.
\ee
One can check that the conservation condition  for ${\cal T}^\mu_\nu$\,,
\be
\dot{{\cal T}^0_0}+3\,\frac{\dot{\alpha}}{\alpha}\,({\cal T}^0_0-{\cal T}^r_r)=0\,,
\ee 
gives exactly the same equation as \eqref{T}, which shows  once again that 
${\cal T}^\mu_\nu$ is identically conserved if ${T}^\mu_\nu$  is conserved. 

The $G^0_0$ equation explicitly reads  
\bea                                \label{q1}
3\,\frac{\dot{\Q}^2+k\,\Q^2}{\Q^4}&=&
m^2\left(
4c_3+c_4-6+\frac{3\beta(3-3c_3-c_4)  }{\Q}
+\frac{3\beta^2(c_4+2c_3-1)}{\Q^2}
-\frac{\beta^3(c_3+c_4)}{\Q^3}
  \right)
+\rho\notag \\
&= &m^2T^0_0+\rho\,,
\eea
while the conservation condition 
\be                          \label{q2}
\{(3c_3+c_4-3)\Q^2+2(1-c_4-2c_3)\Q\beta+(c_3+c_4)\beta^2\}
(\alpha\dot{\Q}-\Q\dot{\beta})=0,
\ee
and the ${\cal G}^0_0$ equation 
\bea                                \label{q3}
3\,\frac{\dot{\beta}^2+k\,\alpha^2}{\alpha^2\beta^2}&=&
\eta m^2\left(
c_4-\frac{3(c_3+c_4)\Q}{\beta}
+\frac{3(c_4+2c_3-1)\Q^2  }{\beta^2}
+\frac{(3-3c_3-c_4)\Q^3  }{\beta^3}
  \right)\notag \\
&= &\eta m^2{\cal T}^0_0\,.
\eea
Let us set the second factor in \eqref{q2} to zero,
\be                       \label{alpha}
\alpha=\frac{\Q\dot{\beta}}{\dot{\Q}}\,,
\ee
thereby solving 
the conservation condition. Setting $\beta(t)=\sigma(t)\Q(t)$
Eq.\eqref{q1} reduces to 
\bea                                \label{q1a}
3\,\frac{\dot{\Q}^2+k\,\Q^2}{\Q^4}&=&
m^2(1-\sigma)\left(
(c_3+c_4)\sigma^2+(3-5c_3-2c_4)\sigma+4c_3+c_4-6)
  \right)
+\rho\notag \\
&=&m^2T^0_0+\rho
\equiv \rho_\ast(\sigma),
\eea
while Eq.\eqref{q3} becomes  
\bea                                \label{q3a}
3\,\frac{\dot{\Q}^2+k\,\Q^2}{\Q^4}&=&
\eta m^2\frac{\sigma-1}{\sigma}\,(c_4\sigma^2-(3c_3+2c_4)\sigma+c_4+3c_3-3)\notag \\
&=&
\eta m^2\sigma^2 {\cal T}^0_0
\equiv \rho_\ast(\sigma).
\eea
We see that the sources of the two metrics are proportional,
\be
m^2T^0_0+\rho=\eta m^2\sigma^2 {\cal T}^0_0,
\ee
where $\sigma=\sigma(\rho)$ fulfills the algebraic equation obtained by 
taking the difference of \eqref{q1a} and \eqref{q3a},
\bea                       \label{algebr}
(c_3+c_4)\sigma^3+(3+\eta c_4-6c_3-3c_4)\sigma^2
+(-9-3\eta c_3-3\eta c_4+9c_3+3c_4)\sigma \notag \\
+\frac{\eta(3-3c_3-c_4)}{\sigma}=
c_4 -6\eta c_3+3\eta-3\eta c_4+4c_3-6  +\frac{\rho}{m^2}
\,. 
\eea
Since $\rho=\rho(\Q)$
in view of the conservation condition \eqref{fluid},
one therefore obtains $\sigma=\sigma(\Q)$.  Injecting this to the right-hand side
of  Eq.\eqref{q3a} (or \eqref{q1a}) gives the source term $\rho_\ast(\Q)$, 
so that the solution $\Q(t)$ can be determined.  

Let us study roots of the quartic equation \eqref{algebr}, first when $c_3+c_4\neq 0$. 
For $\rho=0$ there are generically two real roots, one of which 
is $\sigma=1$ with $\rho_\ast(\sigma)=0$, 
but depending on the parameter 
values there could be altogether four real roots.   
For example, for $\eta=1$, $c_3=-1$ and $c_4=4$ there are four roots 
$\sigma=-0.93,0.56,1,2.19$ with  
$\rho_\ast(\sigma)/m^2=48.78,-0.25,0,-2.42$, respectively.

For non-zero $\rho$  
there generically remain only two real roots, since the other two, if exist, 
merge to each other and disappear when $\rho$ increases. 
When $\rho\to \infty$, one of the two  remaining roots is defined by
\be                              \label{phys}
\frac{\eta(3-3c_3-c_4)}{\sigma}\approx \frac{\rho}{m^2}\,,
\ee 
and the second one is
\bea                       \label{exotic}
(c_3+c_4)\sigma^3\approx \frac{\rho}{m^2}.
\eea
We shall say that the root \eqref{phys} belongs to the physical branch,
since $\sigma$ is small and one can see from \eqref{q1a} that $T^0_0=O(1)$ and
$m^2|T^0_0|\ll \rho$ because $m$ is small, so that $\rho_\ast(\rho)=\rho+O(m^2)$.
 This is physically expected, since 
the  graviton mass contribution to the total energy density 
should normally be small if the matter density is large.   
At the same time, these natural expectations do not apply to the root \eqref{exotic}, 
since $\sigma$ is then large and 
\be
m^2T^0_0=\rho_\ast-\rho=-\rho+O(\rho^{2/3}),
\ee
so that the contribution of the graviton mass to the energy is as large 
as the matter contribution, and the two actually cancel each other, 
up to subleading terms. 
The resulting energy density 
\be
\rho_\ast(\rho)=m^2T^0_0+\rho=c_4\eta m^2\sigma^2+O(\sigma)=
\frac{\eta c_4 m^{2/3}}{|c_3+c_4|^{2/3}}\,\rho^{2/3}+O(\rho^{1/3})
\ee
can even be negative, depending on the sign of $c_4$. We therefore say 
that the root \eqref{exotic} belongs to the exotic branch. 

Both the physical and exotic branches $\rho_\ast(\rho)$ extend from large to 
small values of  $\rho$, so that they  describe the decrease of $\rho$ 
during the universe expansion. 
When the universe is large and 
$\rho\to 0$, the total energy $\rho_\ast$ approaches a constant value
that can be positive or negative or zero, depending on the 
parameter values. For the physical branch $\rho_\ast$ is 
always positive and tends to zero as $\rho\to 0$ 
if  $3-3c_3-c_4>0$, while 
for $3-3c_3-c_4<0$ it approaches a positive value 
(for example, $\rho_\ast\to 15.79$ 
for  $c_3=-1$, $c_4=4$, $\eta=1$). 
For the exotic branch $\rho_\ast$  
is positive/negative at large $\rho$  
if $c_4$ is positive/negative,
respectively, but it seems to always approach 
a non-zero {negative} value when $\rho\to 0$ (if $\eta>0$).  

\begin{figure}[h]
\hbox to \linewidth{ \hss
	
	\psfrag{x}{$\ln(1/\rho)$}
	\psfrag{y}{}
	\psfrag{p}{$\rho_\ast^{\rm phys}$}
	\psfrag{e}{$\rho_\ast^{\rm exotic}$}
	\psfrag{pp}{$\rho/\rho_\ast^{\rm phys}$}
	\psfrag{ee}{$\rho/\rho_\ast^{\rm exotic}$}
	\resizebox{8cm}{5cm}{\includegraphics{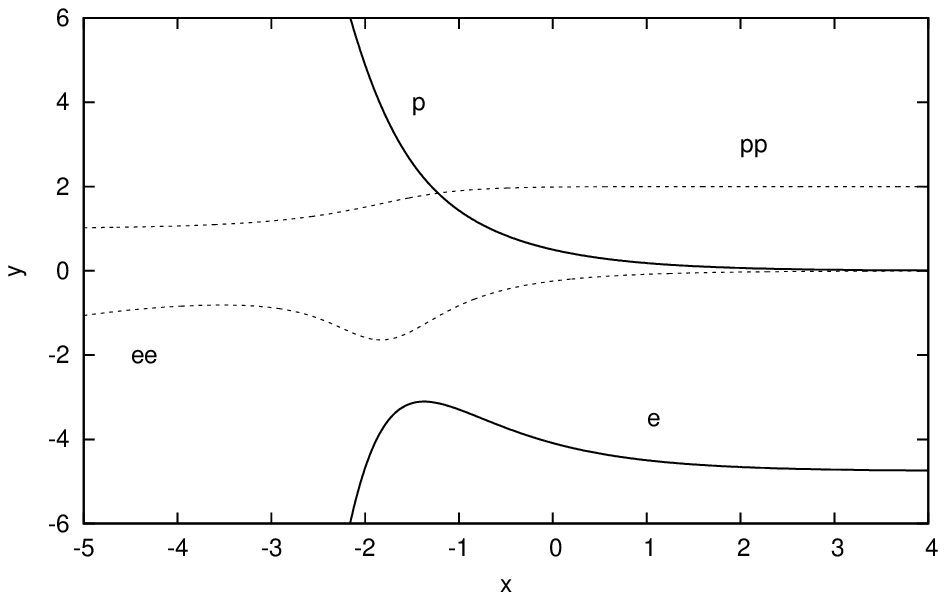}}
\hspace{1mm}
	\resizebox{8cm}{5cm}{\includegraphics{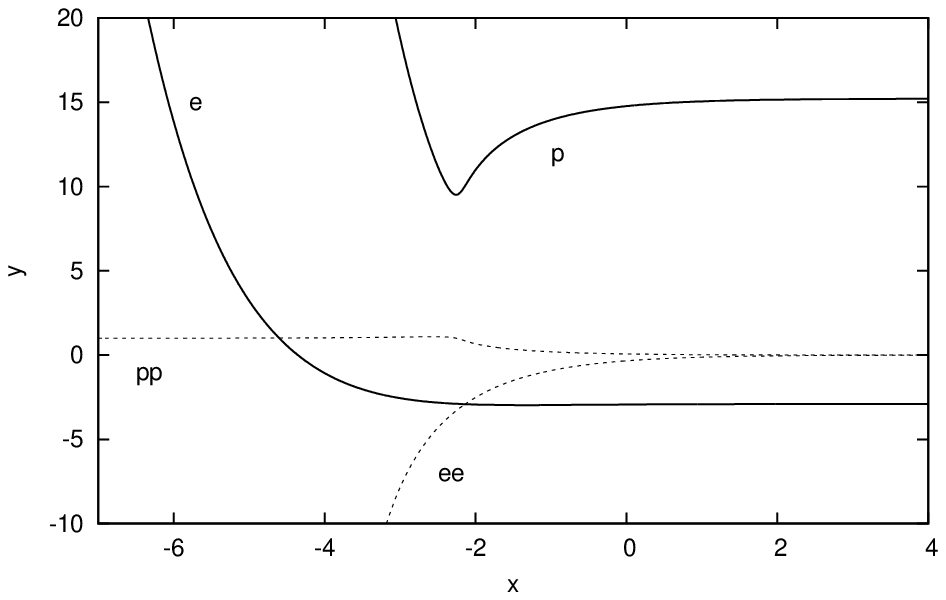}}
	
\hspace{1mm}
\hss}
\label{fig1}
\caption{{\protect\small The total energy density $\rho_\ast=m^2T^0_0+\rho$ 
versus the matter energy density $\rho$ for the physical and exotic branches 
for $c_3=0.9$, $c_4=-1$, $\eta=1$ (left) 
and $c_3=0.9$, $c_4=1$, $\eta=1$ (right). }}%
\end{figure}

The described above different type behaviour of $\rho_\ast(\rho)$ 
can be seen by solving the algebraic equation \eqref{algebr} numerically 
for different parameter values, as shown in  Fig.1.  
One more type of solutions, shown in Fig.2 (left panel),
is obtained by changing the sign of $\eta$, in which case the sign of $\rho_\ast$ along
the exotic branch changes from negative to positive values as $\rho$ decreases.

If $c_3+c_4=0$ then the coefficient in front of the highest power in Eq.\eqref{algebr}
vanishes, so that there remain three roots. One finds in this case 
three different branches $\rho_\ast(\rho)$, these are the physical branch \eqref{phys} and two 
exotic branches that start at large $\rho$ when  one replaces \eqref{exotic} by 
\be
(3+\eta c_4+3 c_4)\sigma^2=\frac{\rho}{m^2}\,,
\ee
since there are two possibilities to choose the sign of $\sigma$  when one takes 
the square root (these branches are called in Fig.2 exotic$+$ and exotic$-$).  
If $\rho$ is large, then 
the energy for both exotic branches is the same up to
subleading terms,
\be
\rho_\ast(\rho)=m^2T^0_0+\rho=c_4\eta m^2\sigma^2+O(\sigma)=
\frac{\eta c_4 }{3+\eta c_4+3 c_4 }\,\rho+O(\rho^{1/2}).
\ee
The behaviour of $\rho_\ast(\rho)$ in the whole range of $\rho$ is shown in Fig.2 (right panel).

\begin{figure}[th]
\hbox to \linewidth{ \hss
	
	\psfrag{x}{$\ln(1/\rho)$}
	\psfrag{y}{}

	\psfrag{p}{$\rho_\ast^{\rm phys}$}
	\psfrag{e}{$\rho_\ast^{\rm exotic}$}
	\psfrag{ph}{$\rho_\ast^{\rm phys}$}
	\psfrag{e+}{$\rho_\ast^{\rm exotic+}$}
	\psfrag{e-}{$\rho_\ast^{\rm exotic-}$}
	\resizebox{8cm}{5cm}{\includegraphics{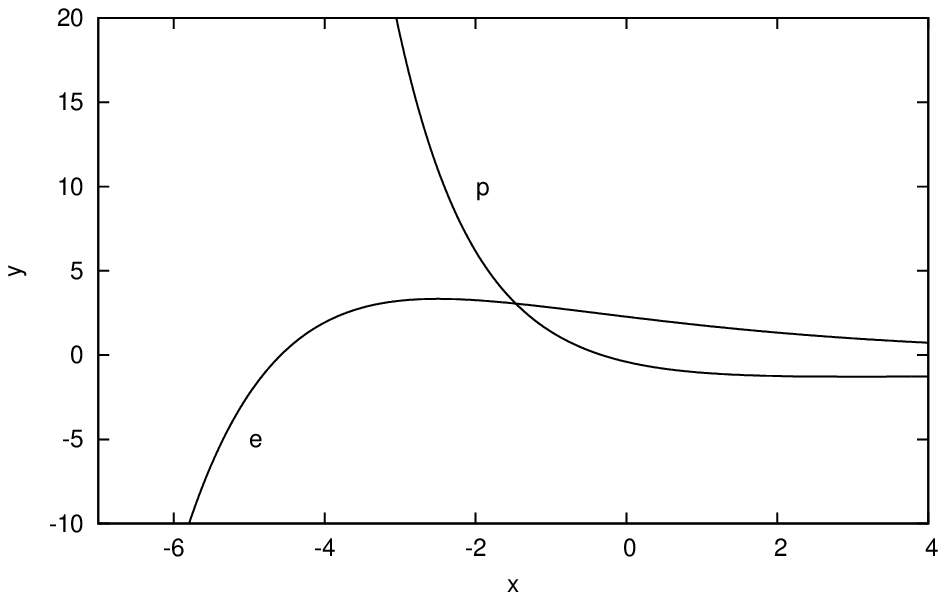}}
\hspace{1mm}
	\resizebox{8cm}{5cm}{\includegraphics{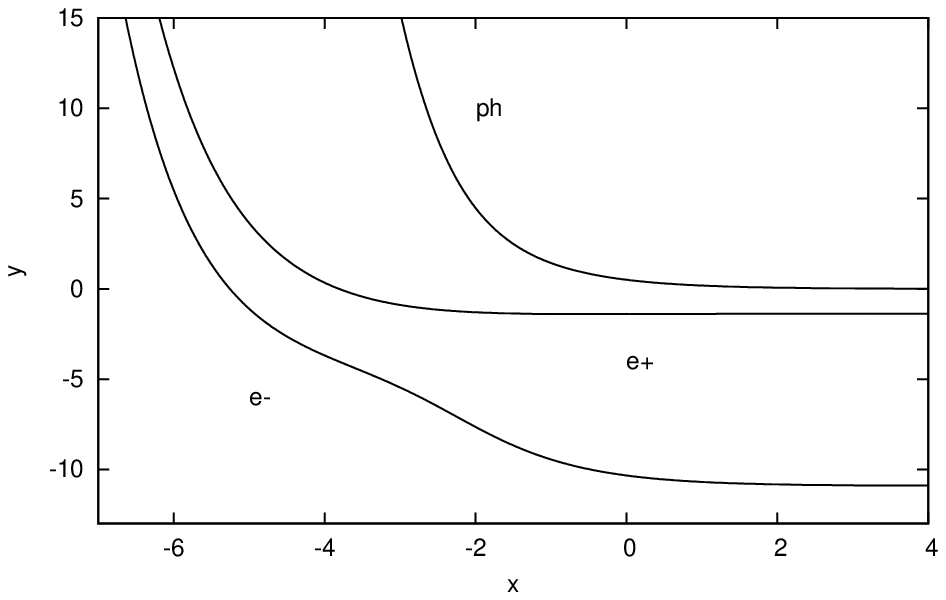}}
	
\hspace{1mm}
\hss}
\label{fig2}
\caption{{\protect\small 
The total energy density $\rho_\ast$ 
versus $\rho$ for the physical and exotic branches 
 for
$c_3=c_4=1$, $\eta=-1$ (left) and 
$c_4=-c_3=1$, $\eta=1$ (right).}}%
\end{figure}

Now that we have determined $\rho_\ast(\rho)$, we can proceed to find solutions 
for the scale factor $\Q(t)$. Introducing the physical time $d\tau=\Q(t)dt$, the 
Einstein equation \eqref{q3a} becomes 
\be                                  \label{energy} 
\left(\frac{d\Q}{d\tau} \right)^2-\frac{\Q^2}{3}\,\rho_\ast(\rho)=-k\,,
\ee 
which describes a `particle' with the total energy $-k=0,\pm 1$ moving in the potential 
\be                      \label{U}
{\rm U}(\Q)=-\frac{\Q^2}{3}\,\rho_\ast(\rho).
\ee 
Assuming a specific equation of state $P=P(\rho)$ for the matter, the conservation
condition \eqref{fluid} gives $\rho(\Q)$, which allows us to compute ${\rm U}(\Q)$. 
In Fig.3 we show ${\rm U}(\Q)$ computed with the ultra-relativistic equation of state, 
\be                                         \label{state}
\rho=3P~~~~~\Rightarrow~~~~~\rho(\Q)=\frac{\rho_0}{\Q^4}. 
\ee
Solutions of Eq.\eqref{energy} for a given $k$ 
correspond to the regions of $\Q$ where ${\rm U}(\Q)\leq -k$. 
Inspecting the ${\rm U}(\Q)$ curves  in Fig.3 reveals then five different cases, 
of which the first two (in the left panel) 
correspond to the physical branches.

Type I solution correspond to the physical branches with $\rho_\ast(\rho)\to 0$ for $\rho\to 0$, 
in which case ${\rm U}(\Q)$ is negative and tends to zero as $\Q\to\infty$. The scale factor
$\Q(\tau )$ behaves qualitatively in the same way as in the matter dominated universe:
it ranges in the finite limits in the spatially closed case $k=1$, it linearly grows with $\tau$
for $k=-1$, and it increases as $\sqrt{\tau}$ for $k=0$. 

Type II solutions correspond to the physical branches with $\rho_\ast(\rho)\to \rho_\ast(0)>0$ (as 
for example in the right part of Fig.1). 
At early times they coincide with 
the ordinary matter-dominated cosmologies, since for the physical branches 
one has $\rho_\ast\approx \rho$ if $\rho$ is large. However, for large $\Q$ one has 
${\rm U}(\Q)= -\rho_\ast(0)\,\Q^2/3$ and $d\Q/d\tau\sim \Q$, so that   
at late times solutions with $k=0,-1$ enter the phase
of accelerated expansion. For $k=1$ the things are slightly more subtle. 
The three curves IIa, IIb and IIc in Fig.3 correspond to different
choices of the integration constant $\rho_0$ in \eqref{state}. If $\rho_0$ is large 
(curve IIc) then ${\rm U}(\Q)<-1$ and the solutions are similar to those with $k=0,-1$. 
If 
$\rho_0$ is small then 
the potential can exceed the value $-1$ (curve IIa). 
Then there is a solution for which the `particle' rests on the left
of the reflection point $A$ (see Fig.3) close to the cosmological singularity $\Q=0$, but 
there is also the solution that stays on the right 
of the reflection point $B$ (see Fig.3), it never approaches  
singularity and shows acceleration  at large $\Q$.  

\begin{figure}[th]
\hbox to \linewidth{ \hss
	
	\psfrag{x}{$\Q$}
	\psfrag{y}{${\rm U}(\Q)/m^2$}
	\psfrag{I}{I}
	\psfrag{IIa}{IIa}
	\psfrag{IIb}{IIb}
	\psfrag{IIc}{IIc}
	\psfrag{IIIa}{III}
	\psfrag{IIIb}{III}
	\psfrag{IIIc}{III}
	\psfrag{IVa}{IV}
	\psfrag{IVb}{V}

	\psfrag{A}{\tiny A}
	\psfrag{B}{\tiny B}
	\psfrag{C}{\tiny C}
	\psfrag{D}{\tiny D}

	\resizebox{8cm}{5cm}{\includegraphics{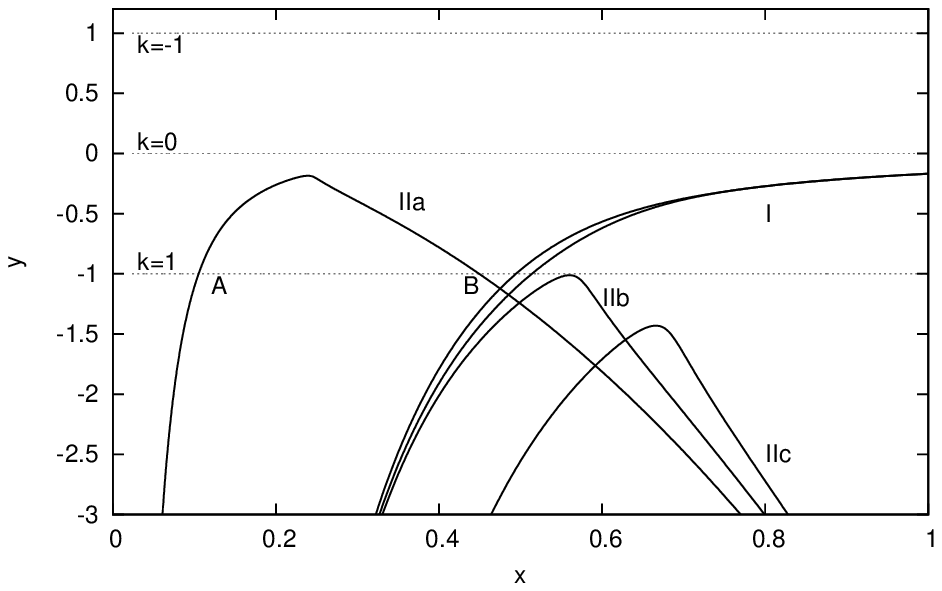}}
\hspace{1mm}
	\resizebox{8cm}{5cm}{\includegraphics{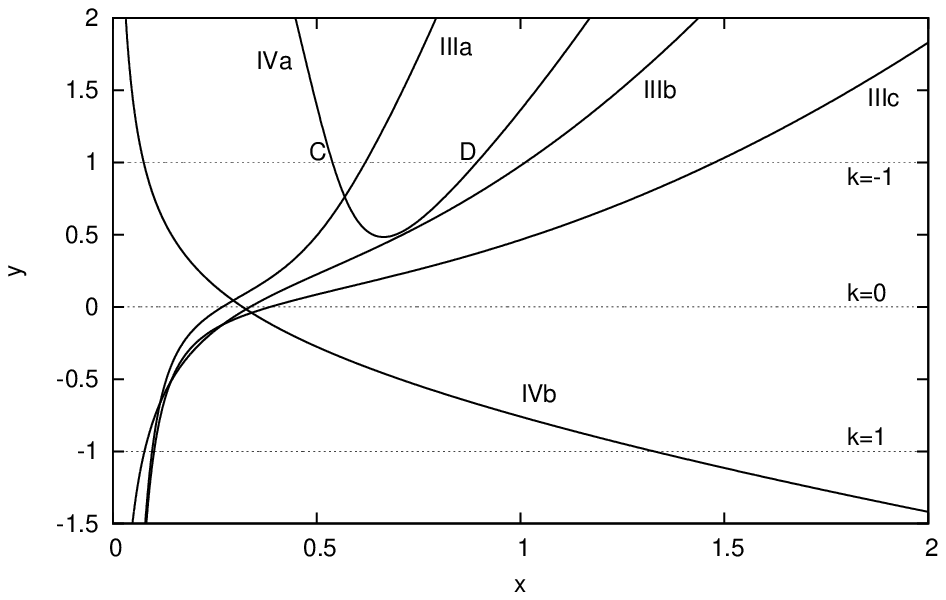}}
	
\hspace{1mm}
\hss}
\label{fig3}
\caption{{\protect\small 
The effective potential ${\rm U}(\Q)$ \eqref{U} for the 
physical (left) and exotic (right) branches  
for solutions shown in Fig.1 and Fig.2.}}%
\end{figure}

The exotic solutions are shown in the right part of Fig.3. Type III corresponds
to exotic branches for which $\rho_\ast(\rho)$ is positive at large $\rho$ but becomes
negative when $\rho$ is small. The potential ${\rm U}(\Q)$ then grows from minus to plus infinity 
and the `particle' is 
always confined to the region close to singularity. 
Type IV corresponds to the exotic branches for which $\rho_\ast$ is always negative,
so that ${\rm U}(\Q)$ is unbounded from above but has a positive minimum. 
Solutions can exist only for $k=-1$ and describe oscillations
in the potential well between the two reflection points ($C,D$ in Fig.3).   
Finally, type V corresponds to the exotic branch for $\eta<0$ for which $\rho_\ast$
changes from negative to positive values as $\rho$ decreases. The potential ${\rm U}(\Q)$ is then 
monotonically decreasing (see Fig.3) and the solutions always stay away from singularity 
and show the self-acceleration at large $\Q$. 

Summarizing, only types II and V show self-accelerating solutions. Type II solutions arise in theories
with $c_4>0$ and $c_3+c_4\neq 0$, they evolve as the matter-dominated universe at early times, 
but enter the accelerated phase at late times. Type V solutions show late time acceleration,
while at early times they are regular, being repelled from the singularity by the negative total energy
$\rho_\ast$. We notice, however, that such solutions require the second gravitational 
coupling constant to be negative.

\section{Non-accelerating solutions  \label{non}}
Let us now return to the conservation equation \eqref{q2} and try to fulfill it 
by setting to zero its first factor and not the second one. We therefore abandon 
the condition \eqref{alpha}, but require instead that $\beta(t)=\sigma\Q(t)$ where 
$\sigma$ is
a constant. Then Eq.\eqref{q2} will be satisfied if 
\be                          \label{q22}
(c_3+c_4)\sigma^2+2(1-c_4-2c_3)\sigma+3c_3+c_4-3=0,
\ee
Eq.\eqref{q1} then reduces to 
\bea                                \label{q11}
3\,\frac{\dot{\Q}^2+k\,\Q^2}{\Q^4}&=&
m^2\left(
4c_3+c_4-6+{3\sigma(3-3c_3-c_4)  }
+{3\sigma^2(c_4+2c_3-1)}
-{\sigma^3(c_3+c_4)}
  \right)
+\rho               \notag \\
&\equiv&\Lambda(\sigma)+\rho\,,
\eea
while Eq.\eqref{q3} becomes 
\be                                \label{q33}
3\,\frac{\dot{\Q}^2}{\alpha^2 \Q^2}+3\,\frac{k}{\sigma^2\Q^2}=
\eta m^2\left(
c_4-\frac{3(c_3+c_4)}{\sigma}
+\frac{3(c_4+2c_3-1)}{\sigma^2}
+\frac{(3-3c_3-c_4) }{\sigma^3}
  \right)=\eta m^2{\cal T}^0_0\,.
\ee
Combining \eqref{q11} and \eqref{q33} one obtains 
\be                                       \label{aa1}
\alpha^2=\sigma^2\Q^2\frac{(\Lambda(\sigma)+\rho )\Q^2-3k}{\eta m^2\sigma^2\Q^2{\cal T}^0_0-3k}\,.
\ee
The quadratic equation \eqref{q22} has two roots,
\be                        \label{rootsig}
\sigma=\frac{2c_3+c_4-1\pm\sqrt{c_3(c_3-1)+c_4+1}}{c_3+c_4},
\ee
and the value of the cosmological constant $\Lambda(\sigma)$ in Eq.\eqref{q11}
is positive 
for one of them and negative for the other one.  
There is, however, an additional condition, since $\alpha^2$ 
must be positive, and as the numerator  in \eqref{aa1} 
is positive because 
$(\Lambda(\sigma)+\rho )\Q^2-3k=3(\dot{\Q}/\Q)^2>0$,  it follows that ${\cal T}^0_0$ should 
be positive, since otherwise the denominator will become negative for large $\Q$.  
Now, 
it turns out that if ${\cal T}^0_0>0$ then $\Lambda(\sigma)<0$,
while if $\Lambda(\sigma)>0$ then ${\cal T}^0_0<0$. 
This eliminates solutions with $\Lambda(\sigma)>0$, but  there remain 
solutions with $\Lambda(\sigma)<0$ and ${\cal T}^0_0>0$, which exist if $k=0,-1$.

\section{Limit $\eta\to 0$   \label{eta}}

Let us first consider the solutions with decoupled metrics  
of Sec.\ref{decoupled}. Taking the limit $\eta\to 0$ 
does not affect the physical metric $g_{\mu\nu}$ 
determined by Eqs.\eqref{a0}, \eqref{a1}, \eqref{fluid}.
On the other hand, the metric 
$f_{\mu\nu}$ in \eqref{AdS}  becomes flat, 
since $F\to 1$ when $\eta\to 0$. One can write 
$f_{\mu\nu}=\eta_{AB}\partial_\mu\Phi^A\partial_\nu\Phi^B$ with 
$\Phi^0=T(t,r)$ and 
$\Phi^a=U(t,r)n^a$\, where $U(t,r)=Cr\Q(t)$ and
$n^a=(\sin\vartheta\cos\varphi,\sin\vartheta\sin\varphi,\cos\vartheta)$,
while $T(t,r)$ is obtained by settin $F=1$ in \eqref{TTT1}:
\be                             \label{unitary} 
T(t,r)=-\frac{Cr^2}{2}\,\dot{\Q}~~{\rm if}~~k=0;~~~~~~~
T(t,r)=kC\dot{\Q}\sqrt{1-kr^2}~~{\rm if}~~k=\pm 1.
\ee
Equation \eqref{a0},\eqref{a1},\eqref{fluid},\eqref{unitary} 
exactly agree with Eqs.(16)--(18) obtained in  \cite{CV} in the RGT limit.   
We therefore conclude that the solutions with decoupled metrics have the counterparts 
in the RGT theory, to which they approach when $\eta\to 0$.

Let us now consider the generic solutions of Sec.\ref{generic}. A direct inspection shows that 
$f_{\mu\nu}$ does not necessarily become flat when $\eta\to 0$, 
because  the source term  $\eta m^2{\cal T}^0_0$ 
in Eq.\eqref{q3}
does not then vanish, neither does the source for the physical metric 
$\rho_\ast=\eta m^2\sigma^2 {\cal T}^0_0$.  
To understand how this is possible, we notice that for the physical branches  
$\sigma$ is small when 
$\rho$ is large, because $\eta/\sigma\sim \rho/m^2$ (see Eq.\eqref{phys}). 
On the other hand, Eq.\eqref{q3a} shows that when $\sigma$ is small then   
$\rho_\ast\sim \eta m^2/\sigma=\rho+\ldots$ for any $\eta$. 
As a result, the effective potential ${\rm U}=-\Q^2\rho_\ast/3$ does not vanish in the region where
$\Q$ is small but approaches a non-trivial limit as $\eta\to 0$ (see Fig.4). 

\begin{figure}[th]
\hbox to \linewidth{ \hss
	
	\psfrag{x}{$\Q$}
	\psfrag{y}{${\rm U}(\Q)/m^2$}
	\psfrag{nu=1}{$\eta=1$}
        \psfrag{nu=0.1}{$\eta=0.1$}
	\psfrag{nu=0.01}{$\eta=0.01$}
	\psfrag{nu=0.001}{$\eta=0.001$}
	\psfrag{nu=0.01,0.001}{$\eta=0.01,0.001$}
	\resizebox{8cm}{5cm}{\includegraphics{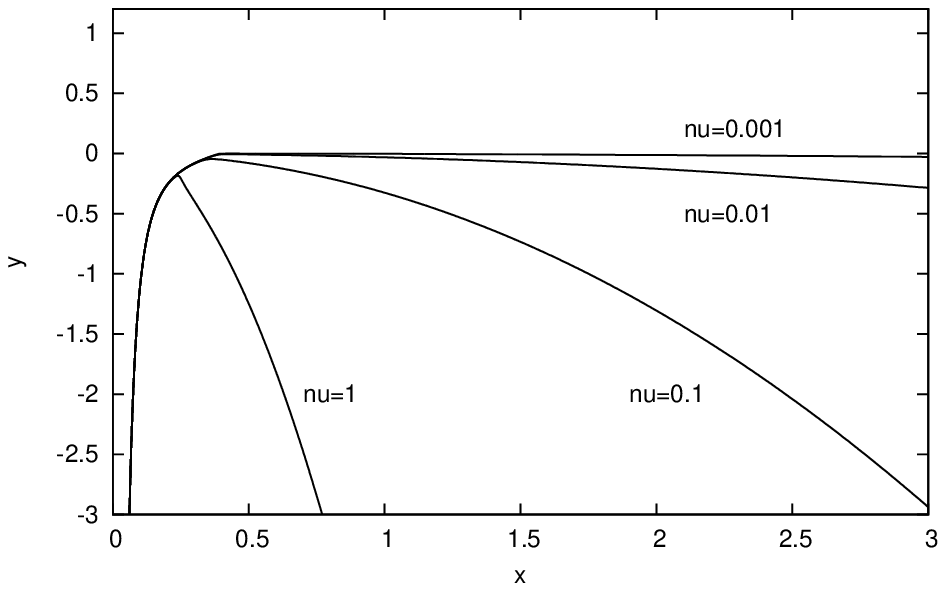}}
\hspace{1mm}
	\resizebox{8cm}{5cm}{\includegraphics{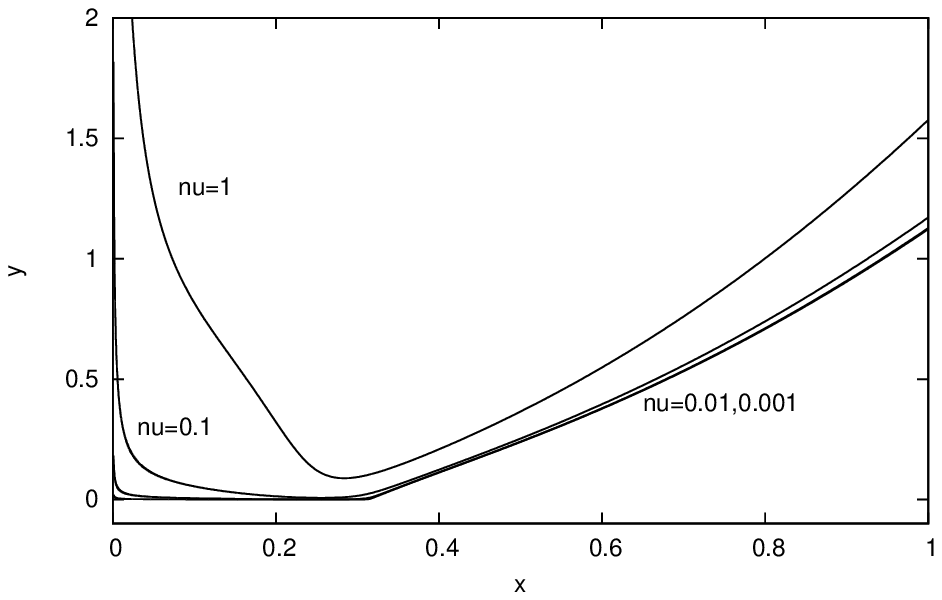}}
	
\hspace{1mm}
\hss}
\label{fig4}
\caption{{\protect\small 
The effective potential ${\rm U}(\Q)$ \eqref{U} for the 
physical solutions with $c_3=0.9$, $c_4=1$ (left) and exotic solutions 
  with $c_3=0.9$, $c_4=-1$ (right) for several values of $\eta$. }}%
\end{figure}

For some exotic branches  
$\sigma$ becomes very small for small $\eta$ when $\rho\to 0$,  in which case 
one finds from  \eqref{algebr} $\sigma\approx \eta(3-3c_3-c_4)/(c_4+4c_3-6)$ so that 
 $\rho_\ast\sim \eta/\sigma$ is independent of $\eta$ and the effective potential $U(\Q)$ does not
vanish at large $\Q$. For other 
exotic branches $\sigma$ never approaches zero, in which case $\rho_\ast\to 0$ as $\eta\to 0$,
therefore
both $g_{\mu\nu}$ and $f_{\mu\nu}$  become flat. 
The conclusion is that the generic solutions of Sec.\ref{generic} 
do not have 
non-trivial analogs in the RGT limit.

Finally, for the solutions of Sec.\ref{non} the metric $g_{\mu\nu}$
is determined by \eqref{q22},\eqref{q11}
and does not depend on $\eta$. The source term in \eqref{q33} vanishes for $\eta\to 0$
and $f_{\mu\nu}$ becomes flat. 
The limit is possible only for $k=-1$, since $\alpha$ in \eqref{aa1} becomes ill-defined
for $k=0$ if $\eta\to 0$. As a result, the solutions of Sec.\ref{non} 
do have, for $k=-1$, analogs in the RTG limit. 
Moreover, for $\eta=0$ one can choose  $\Lambda(\sigma)>0$ in \eqref{q11}, as this 
no longer contradicts the  positivity of $\alpha^2$ in \eqref{aa1}.
Such solutions were found in \cite{Mukohyama}, but only for 
$\Lambda(\sigma)<0$ they can be extended to $\eta\neq 0$. 

We see that the bimetric theory admits solutions which do not approach for $\eta\to 0$ 
those of the $\eta=0$ theory, and vice versa,  the $\eta=0$ theory 
has solutions which do not generalize 
for $\eta\neq 0$.  One can construct more solutions for $\eta=0$
if we go directly to Eqs.\eqref{q1},\eqref{q2} and require that the metric 
parameterized by the functions
$\alpha,\beta$ (with $U=r\beta(t)$),
\be
f_{\mu\nu}dx^\mu dx^\nu=\alpha(t)^2dt^2-\frac{\beta(t)^2}{1-kr^2}\,dr^2
-U^2(d\vartheta^2+\sin^2\vartheta d\varphi^2),
\ee
be flat. It will be flat if one finds 
$T(t,r)$ such that 
\be                                  \label{flat}
dT^2-dU^2=\alpha(t)^2dt^2-\frac{\beta(t)^2}{1-kr^2}\,dr^2\,,
\ee
which 
 is equivalent to three conditions
\bea                              \label{flat1}
{\beta^2}-T^{\prime 2}=\frac{\beta(t)^2}{1-kr^2},~~~~~~~
\dot{T}^2-r^2\dot{\beta}^2=\alpha^2,~~~~~~
\dot{T}T^\prime=r\dot{\beta}\beta.
\eea
One possibility to fulfill these conditions is to set $\alpha=0$, $\beta=C$, and
\be                                              \label{Tsing}
k=0:~~~T=0,~~~~~~~~k=\pm 1:~~~ T=\frac{C}{\sqrt{-k}}\,\sqrt{1-kr^2}\,\,. 
\ee
The conservation condition \eqref{q2} is then fulfilled and one is left with the 
Einstein equation \eqref{q1} where $\beta$ is constant. 
This reproduces the solutions given by Eqs.(19),(20) in \cite{CV}
(the opposite sign convention for $c_3$ is used in \cite{CV}),  they 
exist only in the RGT limit and do not generalize for $\eta\neq 0$.

Another possibility to fulfill \eqref{flat1} is to choose $k=-1$ and set 
\be
T=\sqrt{1+r^2}\,\beta(t),~~~~~~~~\alpha=\dot{\beta}\,.
\ee
With $\beta(t)=\sigma\Q(t)$ the conservation condition \eqref{q2} is fulfilled 
if $\sigma$ is given by \eqref{rootsig}, while $\Q$ is  then determined by \eqref{q11}.   
The solutions were obtained in Ref.\cite{Mukohyama}, they  generalize to $\eta\neq 0$
if only one chooses the root of \eqref{rootsig} for which $\Lambda(\sigma)$ in  \eqref{q11}
is negative. 

Summarizing, among the accelerating solutions of the RGT theory 
only the special solutions \eqref{unitary}  generalize
for $\eta\neq 0$, while among 
accelerating solutions of the bimetric theory 
only solutions with the decoupled metrics of Sec.\ref{decoupled} have 
the RGT limit.

Recently it was claimed in the literature 
that the RTG theory 
does not actually admit homogeneous and isotropic cosmological solutions 
\cite{GAB} (apart from those obtained in the decoupling limit \cite{decouple}). 
At the same time, the 
presented above analysis shows very explicitly 
that such solutions exist, thus confirming the results of  \cite{CV}, \cite{Mukohyama}. 
The negative argument of \cite{GAB} assumes 
that in the unitary gauge, where $\Phi^\mu=x^\mu$ and $f_{\mu\nu}=\eta_{\mu\nu}$, 
the physical metric $g_{\mu\nu}$ is diagonal
(see Eq.(13) in \cite{GAB}). 
However, the two metrics cannot in general be diagonal at the same time.
For example,  $g_{\mu\nu}$ is diagonal in coordinates 
$t,r,\vartheta,\varphi$, but  $f_{\mu\nu}=\eta_{AB}\partial_\mu\Phi^A\partial_\nu\Phi^B$
with $\Phi^A$ defined by formulas around Eq.\eqref{unitary} is not diagonal. 
 For the solutions \eqref{Tsing}
both metrics are diagonal at the same time, but $f_{\mu\nu}$
is degenerate, so that the argument of \cite{GAB} again does not apply.

\section{Summary \label{final}}

We have presented the homogeneous and isotropic cosmological solutions 
within the bimetric generalization of the RGT massive gravity theory. 
These solutions can be spatially open, 
closed, or flat, and at early times they are sourced by the perfect fluid, 
while the graviton mass typically manifests itself 
at late times by giving rise to a cosmological term
whose value is determined 
by the theory parameters $c_3,c_4,\eta$. In addition, there are also 
exotic solutions for which already at  early times, when the matter 
density $\rho$ is high, the contribution of the graviton mass to the energy density
is large and screens that of the matter contribution. 
The total energy $m^2T^0_0+\rho$ can be negative, which can lead to non-singular 
solutions, as in the case of type II solutions with $k=1$ of Sec.\ref{generic}. 
For type V solutions of Sec.\ref{generic} the cosmological singularity is removed
altogether, but this requires the second gravitational coupling to be negative. 

In the limit where the second gravitational coupling tends to zero the generic 
solutions of Sec.\ref{generic} do not reduce to solutions of the RGT theory, 
since both metrics  remain then curved. However, the special solutions with 
decoupled metrics do have the non-trivial RGT limit.

The analysis of stability of our solutions remains an open issue to study. 
Since the graviton contribution to the total energy can be negative and very large 
for the exotic solutions, it is not impossible that the ghost could be still present
in the theory, which may affect the stability.

\renewcommand{\thesection}{APPENDIX}
\section{}
\renewcommand{\theequation}{A.\arabic{equation}}
\setcounter{equation}{0}
\setcounter{subsection}{0}
Here we list the energy momentum tensor components
in the spherically symmetric case. Using the expression \eqref{gamma}
for $\gamma^\mu_{~\nu}$ and  computing $K^\mu_\nu=\delta^\mu_\nu-\gamma^\mu_{~\nu}$
gives the following value of the interaction
Lagrangian \eqref{lagr}: 
\bea
{\cal L}_{\rm int}&=& 6+\frac{ab}{SN}+\frac{c^2}{N^2}
-\frac{3a}{S}-\frac{3b}{N}+\frac{2aU}{SR}+\frac{2bU}{NR}-\frac{6U}{R}
+\frac{U^2}{R^2} \notag \\
&-&c_3\frac{R-U}{R}\left(\frac{2ab}{NS}
-\frac{3b}{N}-\frac{3a}{S}+4+\frac{2c^2}{N^2}
+\frac{Ua}{RS}-\frac{2U}{R}+\frac{bU}{RN} \right)\notag \\
&-&c_4\frac{(R-U)^2}{R^2}\left(1-\frac{a}{S}
-\frac{b}{N}+\frac{ab}{NS}+\frac{c^2}{N^2}  \right),
\eea
while the non-zero components of $\tau^\mu_\nu$ defined by Eq.\eqref{tau} read
\bea
\tau^0_0&=&\frac{ab}{SN}+\frac{c^2}{N^2}-\frac{3a}{S}
+\frac{2aU}{SR}
+c_3\,\frac{R-U}{R}\,\left(\frac{3a}{S}
-\frac{2ab}{SN}-\frac{2c^2}{N^2}-\frac{aU}{SR} \right)\notag \\
&+&c_4\,\frac{(R-U)^2}{R^2}\,
\left(\frac{a}{S}-\frac{ab}{SN}-\frac{c^2}{N^2}  \right), 
\eea
\bea
\tau^r_r&=&\frac{ab}{SN}+\frac{c^2}{N^2}-\frac{3b}{N}+\frac{2bU}{NR}
+c_3\,\frac{R-U}{R}\,\left(
\frac{3b}{N}-\frac{2ab}{SN}
-\frac{2c^2}{N^2}-\frac{bU}{NR}  
\right)\notag \\
&+&c_4\,\frac{(R-U)^2}{R^2}\,
\left(\frac{b}{N} 
-\frac{ab}{SN}
-\frac{c^2}{N^2}    \right),
\eea
\bea
\tau^\vartheta_\vartheta&=&\tau^\varphi_\varphi=
\frac{U}{R}\left(\frac{a}{S}+\frac{b}{N}-3+\frac{U}{R}\right)
+c_3\,\frac{U}{R}\,\left(
3-\frac{2b}{N}
-\frac{2U}{R}+\frac{bU}{NR}
-\frac{2a}{S}+\frac{aU}{SR}+\frac{ab}{SN}
+\frac{c^2}{N^2}  \right) \notag \\
&+&c_4\,\frac{U(R-U)}{R^2}\,
\left(  
1-\frac{a}{S}-\frac{b}{N}+\frac{ab}{SN}+\frac{c^2}{N^2}
\right), 
\eea
\bea
\tau^0_r=\frac{c}{R^2S}
\left( -R\,(3R-2U)+ c_3\,(3R-U)(R-U)
+c_4\,(R-U)^2 \right).
\eea
The components of the two energy-momentum tensor are then simply obtained from  
Eq.\eqref{TTT}, where $\sqrt{-g}/\sqrt{-f}$ is given by \eqref{det}.

\end{document}